\documentclass[apj]{emulateapj}

\usepackage{times}

\shorttitle{Observing Thermal Flare Plasma with EVE}
\shortauthors{Warren, Mariska, \& Doschek}

\begin{document}


\title{Observations of Thermal Flare Plasma with the EUV Variability Experiment}

\author{Harry P. Warren\altaffilmark{1}, John T. Mariska\altaffilmark{2}, and George
  A. Doschek\altaffilmark{1}}

\affil{$^1$Space Science Division, Naval Research Laboratory, Washington, DC 20375}

\affil{$^2$School of Physics, Astronomy, and Computational Sciences, George Mason
  University, 4400 University Drive, Fairfax VA 22030}


\begin{abstract}
  One of the defining characteristics of a solar flare is the impulsive formation of very
  high temperature plasma. The properties of the thermal emission are not well understood,
  however, and the analysis of solar flare observations is often predicated on the
  assumption that the flare plasma is isothermal. The EUV Variability Experiment (EVE) on
  the \textit{Solar Dynamics Observatory} (\textit{SDO}) provides spectrally resolved
  observations of emission lines that span a wide range of temperatures (e.g.,
  \ion{Fe}{15}--\ion{Fe}{24}) and allow for thermal flare plasma to be studied in detail.
  In this paper we describe a method for computing the differential emission measure
  distribution in a flare using EVE observations and apply it to several representative
  events. We find that in all phases of the flare the differential emission measure
  distribution is broad. Comparisons of EVE spectra with calculations based on parameters
  derived from the \textit{GOES} soft X-ray fluxes indicate that the isothermal
  approximation is generally a poor representation of the thermal structure of a flare.
\end{abstract}



 \section{introduction}

 It is widely believed that solar flares are powered by magnetic reconnection
 \cite[e.g.,][]{priest2002}. How this energy is released over short time scales relative
 to magnetic diffusion is not well understood. Since a significant fraction of the energy
 released during a solar flare is ultimately radiated away, fully characterizing thermal
 emission and how it evolves with time is critical for providing strong observational
 constraints on theoretical models. During the past several decades flare observations
 have generally focused on either broad-band, soft X-ray measurements or high spectral
 resolution measurements of individual emission lines. Such observations have been unable
 to determine the full distribution of plasma temperatures in a flare and there have been
 only a few calculations of flare differential emission measure (DEM) distributions that
 extend to temperatures of 5\,MK and lower \citep[e.g.,][]{mctiernan1999,dere1979}.

 The launch of the EUV Variability Experiment (EVE, \citealt{woods2012}) on the
 \textit{Solar Dynamics Observatory} (\textit{SDO}) provides a new opportunity to study
 thermal flare plasma. EVE is a spectral irradiance monitoring instrument that observes
 the full Sun at wavelengths between approximately 65 and 1050\,\AA\ with 1\,\AA\ spectral
 resolution and a 10\,s cadence. The spectral coverage of EVE allows for the observation
 of flare emission lines in the 90--150\,\AA\ wavelength range, which has not been
 observed systematically for many decades \cite[e.g.,][]{kastner1974}. Emission lines in
 this spectral range cover \ion{Fe}{18} to \ion{Fe}{23} and, in combination with
 observations of other emission lines in the EUV such as \ion{Fe}{15} 284.16\,\AA\ and
 \ion{Fe}{24} 192.04\,\AA, provide a complete description of thermal flare plasma at
 temperatures from 2 to 30\,MK. Moreover, the \ion{Fe}{21} lines in this wavelength range
 are sensitive to the electron density \citep{mason1979,milligan2012} and provide
 information on the emission-measure-weighted density in the flare.

 In this paper we describe a method for calculating the differential emission measure
 using EVE observations and illustrate the application of this method using several long
 duration events associated with coronal mass ejections. We focus on two-ribbon, eruptive
 events because the magnetic geometry of the post-eruption arcade, a Y-type current sheet,
 appears to be consistent with the observations and provides perhaps the simplest
 environment in which to study magnetic reconnection in the solar atmosphere. Furthermore,
 two-ribbon flares provide an ideal way to test the hydrodynamics of loop evolution. The
 observations suggest that flare loops are heated impulsively leading to the evaporation
 of material into the corona, and subsequent cooling to lower temperatures. It remains to
 be seen, however, whether 1-D hydrodynamic simulations can accurately model the flow of
 mass and energy through the solar atmosphere in even this very simple case.

 For the events we consider we find a relatively broad DEM during all phases of the flare.
 As expected, the highest temperature emission is observed during the rise phase and at
 the peak of the event. During the decay phase of the flare the magnitude of the emission
 measure decays exponentially but the peak temperature declines very slowly. We also find
 that the shape of the temperature distribution remains relatively constant as the flare
 decays over many hours, suggesting that plasma temperature is relatively insensitive to
 the magnitude of the energy released by the magnetic reconnection process.

 For these observations we also compute the isothermal temperature and emission measure
 from the ratio of the \textit{GOES} soft X-ray fluxes and use these parameters to infer
 the expected EVE spectrum. These comparisons show that the differential emission measure
 distribution reproduces the observations at EUV wavelengths much better than an
 isothermal model. Single temperature fits are often used in the analysis and
 interpretation of solar flare observations \citep[e.g.,][]{sui2003}.

 \section{Observations}

 \begin{figure*}[t!]
 \centerline{\includegraphics[height=7.0in,angle=90]{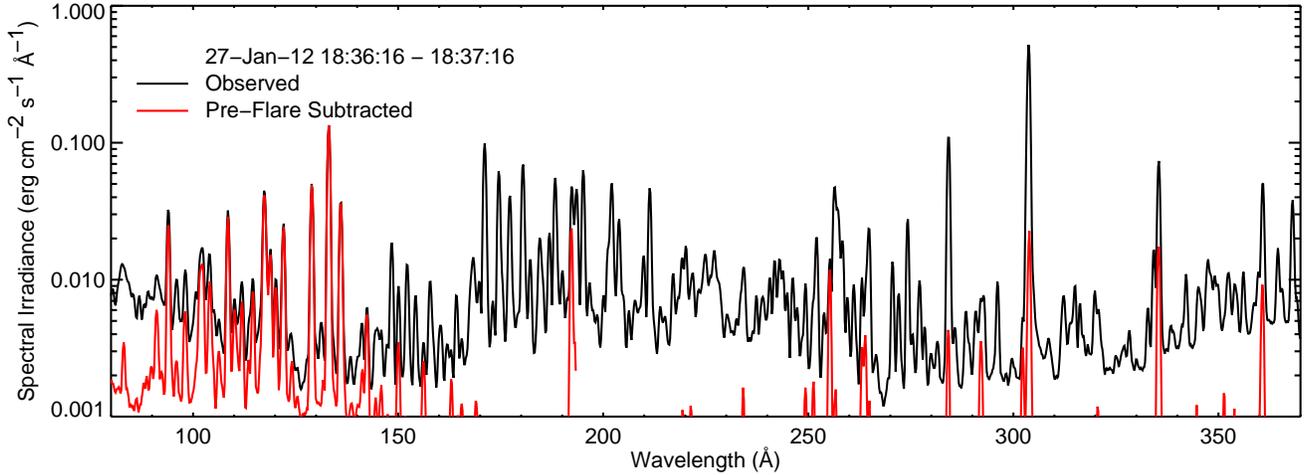}}
 \caption{EVE observations of the solar spectral irradiance near the peak of an X1.7 flare
   that occurred on 2012 January 27. Both the observed and pre-flare subtracted spectra
   are shown. Emission lines from \ion{Fe}{15} to \ion{Fe}{24} are evident in the flare
   spectrum. }
 \label{fig:eve}
 \end{figure*}

 \begin{figure}[t!]
 \centerline{\includegraphics[width=3.5in]{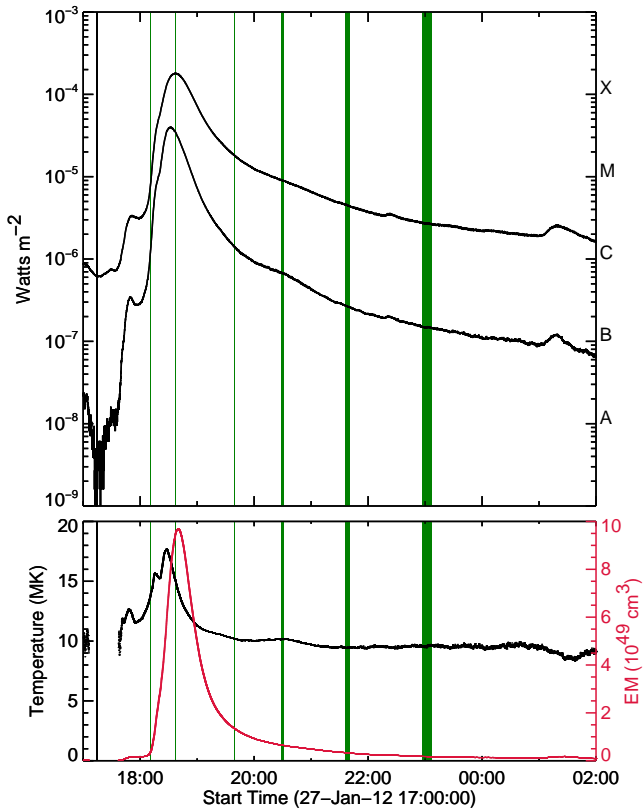}}
 \caption{\textit{GOES} observations of the 2012 January 27 X1.7 flare. The top panel
   shows the \textit{GOES} soft X-ray 0.5--4 and 1--8\,\AA\ fluxes as a function of
   time. The vertical lines indicate intervals selected for determining the background for
   both EVE and \textit{GOES} as well as other times of interest during the rise, peak,
   and decay of the flare. The bottom panel shows the isothermal temperature and emission
   measure derived from the \textit{GOES} fluxes.}
 \label{fig:goes}
 \end{figure}

 EVE is actually a collection of instruments designed to measure the solar irradiance at
 many EUV wavelengths. In this work we will consider observations from the Multiple EUV
 Grating Spectrograph A (MEGS-A), which is a grazing incidence spectrograph that observes
 in the 50 to 370\,\AA\ wavelength range. MEGS-A has a spectral resolution of
 approximately 1\,\AA\ and an observing cadence of 10\,s. For more detail see
 \cite{woods2012}.

 An example EVE spectrum from the peak of the X1.7 flare that occurred on 2012 January 27
 between approximately 17 and 22 UT is shown in Figure~\ref{fig:eve}. We will use this
 event to describe our analysis in detail and then consider other events more
 briefly. Figure~\ref{fig:eve} clearly shows the utility of the MEGS-A wavelength range
 for determining the properties of thermal flare plasma. Observations between 90 and
 150\,\AA, where we find some of the most intense emission lines from
 \ion{Fe}{18}--\ion{Fe}{23}, are a particularly rich source of diagnostics.  Additionally,
 there are the strong \ion{Fe}{24} 192.04 and 255.10\,\AA\ flare lines within the MEGS-A
 wavelength range. Observations of \ion{Fe}{15} 284.16\,\AA\ and \ion{Fe}{16} 335.41\,\AA\
 provide information on lower temperature plasma.

 For this work we also analyze observations from the soft X-ray monitors on the
 \textit{Geostationary Operational Environmental Satellites} (\textit{GOES}), which
 provide spatially integrated fluxes in the 1--8 and 0.5--4\,\AA\ wavelength ranges at a
 2\,s cadence. These bandpasses have contributions from \ion{Fe}{25} and \ion{Fe}{26}
 emission lines. These ions are formed at somewhat higher temperatures than the lines
 found in the MEGS-A wavelength range. Free-free continuum is also important at these
 wavelengths. Additional details on the \textit{GOES} soft X-ray monitors can be found in
 \citet{garcia1994}, \citet{white2005}, and \citet{aschwanden2012}.

 The \textit{GOES} light curves for the 2012 January 27 X1.7 event are shown in
 Figure~\ref{fig:goes}.  We use the \textit{GOES} light curves to identify various times
 of interest during the rise, peak, and decay of the flare. These six times are indicated
 on the light curves shown in Figure~\ref{fig:goes}. For the analysis of both the
 \textit{GOES} and EVE data background subtraction is necessary to isolate the
 contribution of the flare. For this we use a 60\,s interval centered around the lowest
 observed flux in the \textit{GOES} 0.5--4\,\AA\ channel during the 30 minutes preceding
 the peak of the flare. This time interval is also indicated in Figure~\ref{fig:goes}.

 \begin{figure*}[t!]
   \centerline{\includegraphics[width=6.5in]{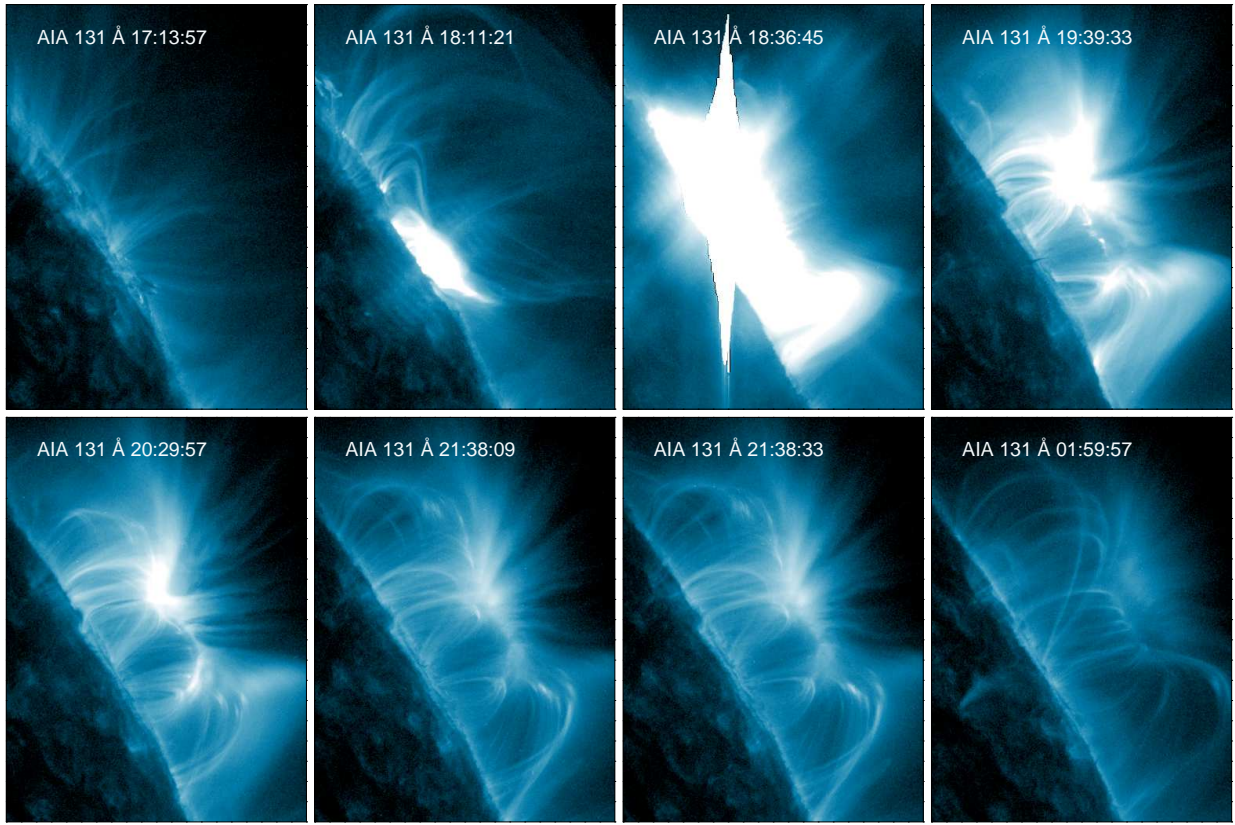}}
   \caption{AIA 131\,\AA\ images from various times during the 27 January 2012 X1.7 flare
     and coronal mass ejection. All of the images shown here have the same logarithmic
     scaling. This AIA channel is sensitive to both high temperature flare plasma from
     \ion{Fe}{21} 128.75\,\AA\ and lower temperature emission from \ion{Fe}{8} emission
     lines.}
   \label{fig:aia}
 \end{figure*}

 Like \textit{GOES}, EVE makes spatially unresolved observations. In contrast to
 \textit{GOES}, the EUV irradiance is generally dominated by non-flare emission (see
 Figure~\ref{fig:eve}). Also, at a spectral resolution of 1\,\AA, the vast majority of the
 flare lines in the EUV are blended with other lines formed at much cooler
 temperatures. To further complicate the analysis many of these lower temperature emission
 lines are unidentified and there is no atomic data for them
 \citep[e.g.,][]{testa2012}. Given these constraints the best strategy is to remove the
 lower temperature emission by subtracting a pre-flare observation from the EVE
 measurements during the event. The primary risk in this approach is that the lower
 temperature emission will also evolve during the event. For example, for eruptive events
 dimming is often observed in emission lines formed around 1\,MK
 \citep[e.g.,][]{gopalswamy1998}, which would lead to an underestimate of flare emission.

 To prepare the EVE observations for analysis we preformed some additional processing to
 the calibrated level2 data.  We first computed time-averaged EVE spectra for each of the
 time intervals indicated in Figure~\ref{fig:goes}. We use the observed standard deviation
 in the irradiance measurements ($\sigma_I$) to estimate the statistical uncertainty in
 each spectral bin, $\sigma_{\bar{I}} = \sigma_I/\sqrt{N}$, where N is the number of
 spectra in the average.  We then subtracted the pre-flare, background spectrum from each
 of these spectra and propagate the errors in the usual way. The statistical uncertainties
 for the background subtracted flare irradiances are generally very small (approximately
 1\% at most wavelengths) and are likely to be dominated by systematic errors in the
 analysis, such as the assumption that the non-flare irradiance is constant during the
 event. 

 Both free-free and free-bound continuum emission has been observed during flares with EVE
 \citep{milligan2012}, but to simplify our analysis we have chosen to remove it. To
 accomplish this we determine the lowest intensity in each 10\,\AA\ wavelength bin and
 subtract this from the observed spectrum.  An example background subtracted spectrum from
 the peak of the X1.7 flare is shown in Figure~\ref{fig:eve}.

 Finally, to provide context for these observations we have investigated the images from
 the AIA instrument on \textit{SDO} \citep{lemen2012}. AIA is a set of multi-layer
 telescopes capable of imaging the full Sun at high spatial resolution (0.6\arcsec\
 pixels) and high cadence (typically 12\,s). Images are available at 94, 131, 171, 193,
 211, 304, and 335\,\AA. AIA images are also available at UV and visible wavelengths, but
 they are not used in this analysis. The AIA 131\,\AA\ includes contributions from
 \ion{Fe}{21} 128.75\,\AA\ and is particularly useful for flare observations. It also
 includes contributions from lower temperature emission lines. Figure~\ref{fig:aia} shows
 several images from the event. An animation of these data clearly shows an eruption
 followed by the formation of a classical post-flare loop arcade.

 \begin{figure*}[t!]
 \centerline{\includegraphics[height=6.5in,angle=90]{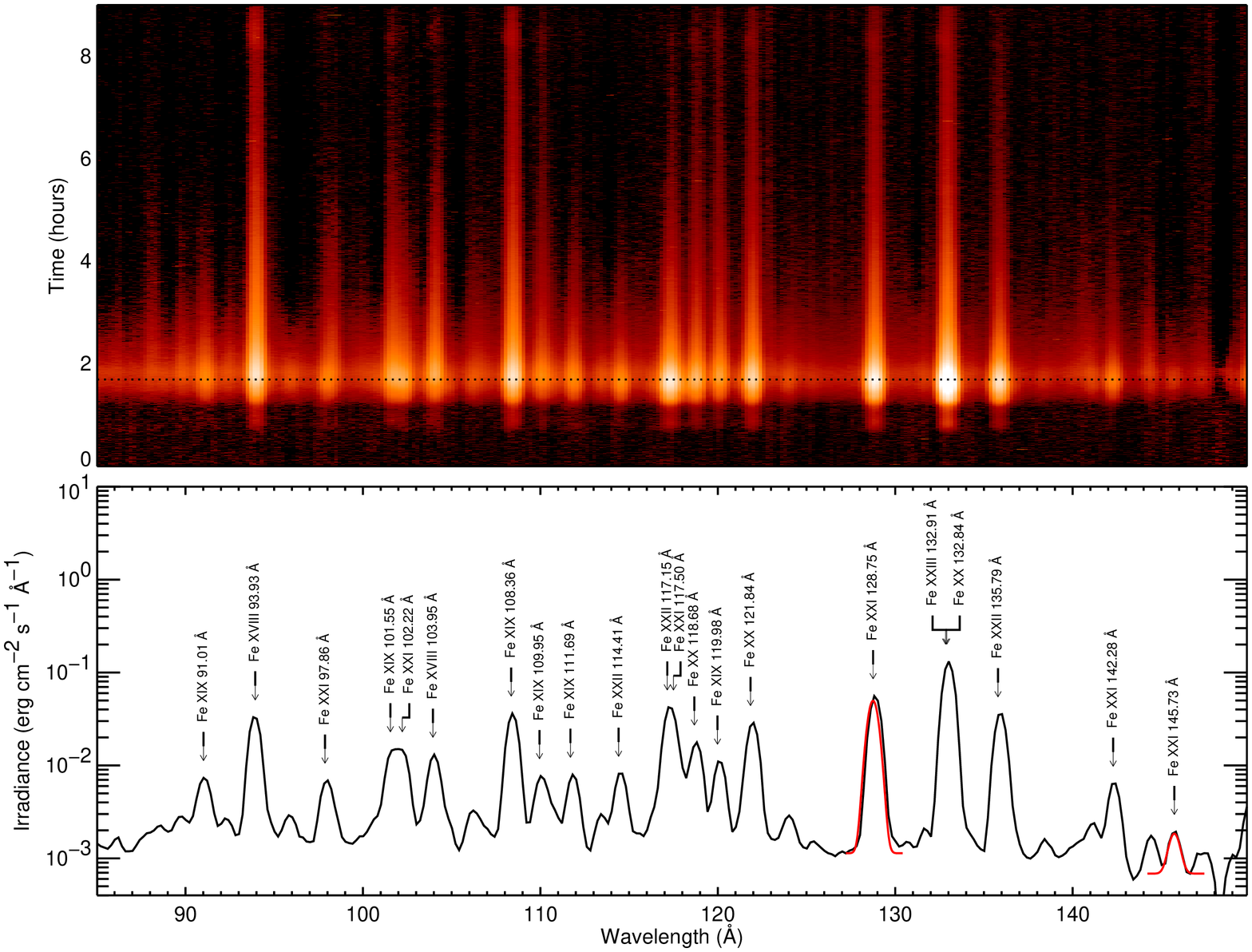}}
 \centerline{\includegraphics[height=6.5in,angle=90]{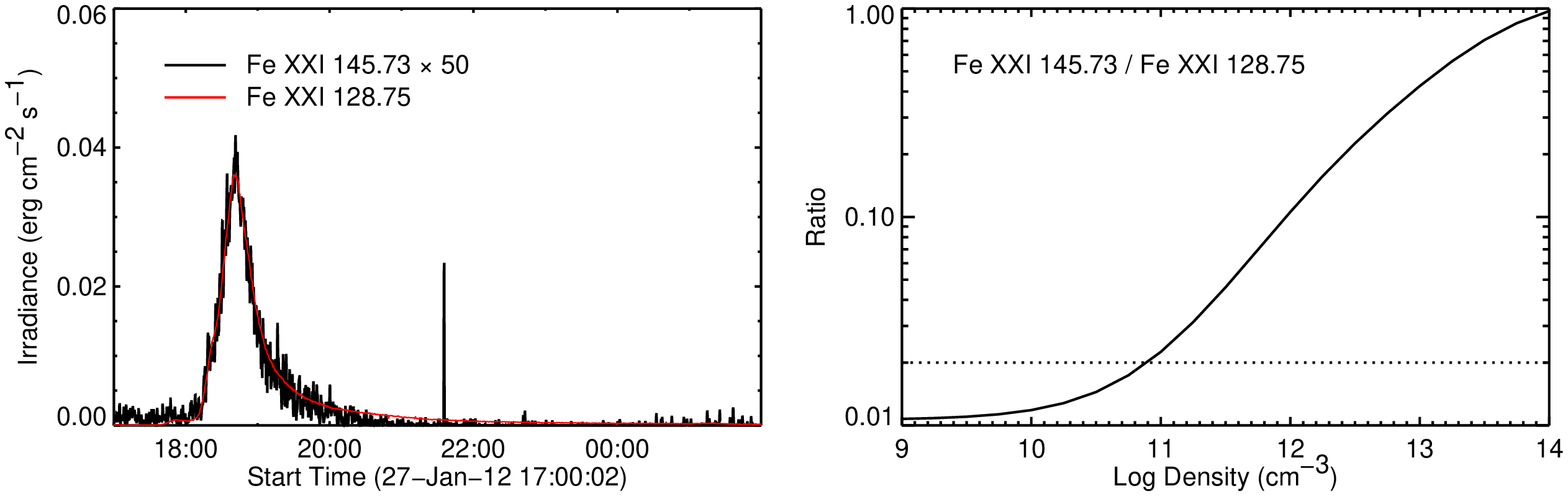}}
 \caption{EVE observations of the 85-150\,\AA\ wavelength range for the X1.7 flare that
   occurred on January 27, 2012. The top panel shows the spectra as a function of time and
   wavelength. The middle panel shows the spectrum at the peak of the event with many of
   the most intense emission lines identified. A pre-flare spectrum has been subtracted
   from the observed irradiance during the flare. The peak intensities above the local
   continuum to estimate the total intensity of \ion{Fe}{21} 128.75 and 145.72\,\AA. The
   bottom panels shown the evolution these lines. During most of the event the line ratio
   is approximately 0.02 corresponding to a density of $10^{11}$\,cm$^{-3}$. This
   theoretical ratio was computed using the CHIANTI database.}
 \label{fig:eve2}
 \end{figure*}

 \section{Flare Density and Emission Measure}

 In this section we will determine the density and temperature evolution for the 2012
 January 27 event. We will consider both the isothermal emission measure model that can be
 derived from the \textit{GOES} observations as well as the DEM model that can be derived
 from EVE. Calculations of plasma emissivities, which form the basis of the temperature
 analysis, require information on the density so we begin with electron densities that can
 be inferred from the \ion{Fe}{21} emission lines observed in the EVE spectra.
 
 \begin{figure*}[t!]
   \centerline{\includegraphics[height=7.0in,angle=90]{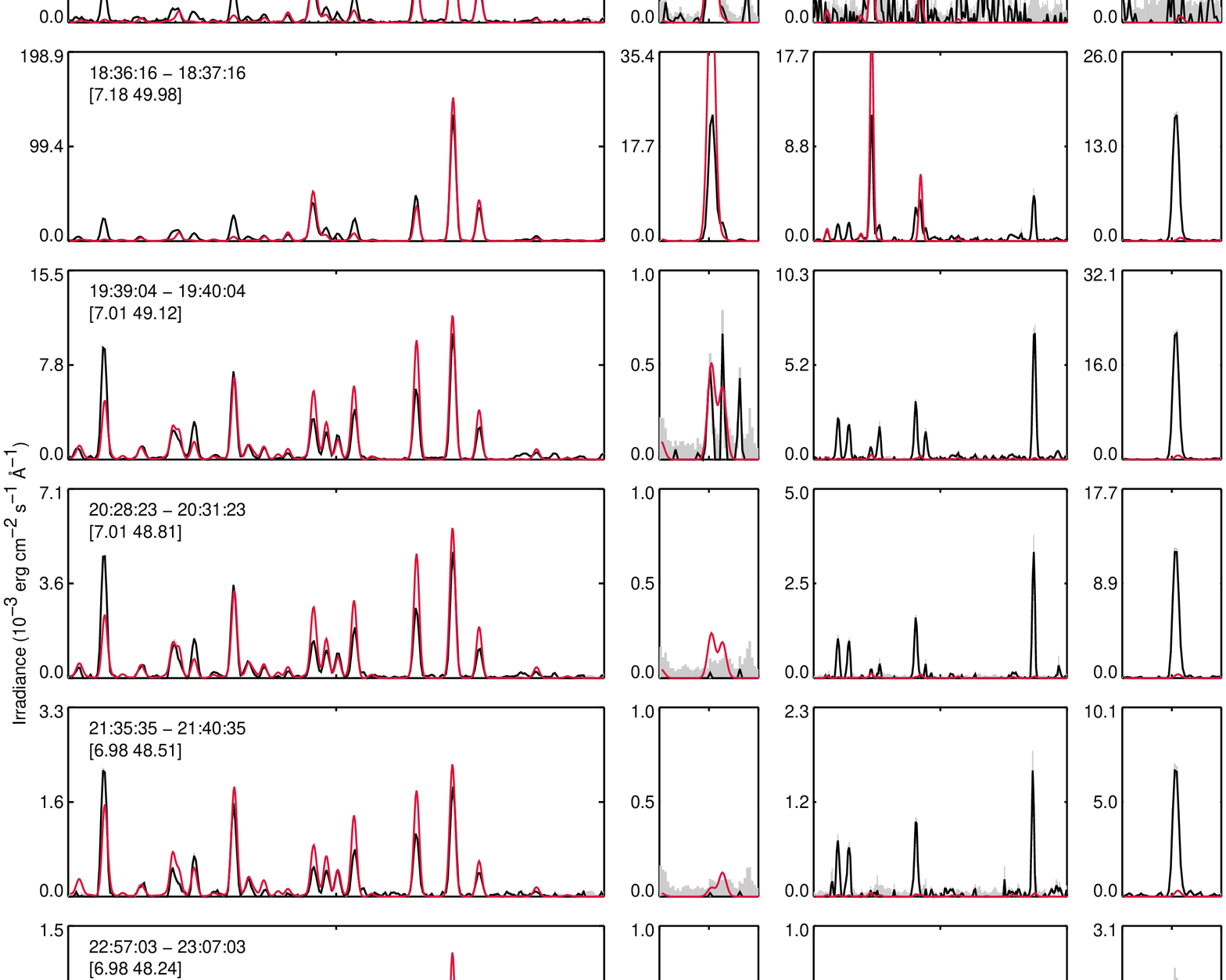}}
   \caption{Comparison of observed EVE flare spectra with spectra inferred from the
     \textit{GOES} isothermal temperature and emission at various time during the 2012
     January 27 event. The grey bands indicate the statistical uncertainty in the
     background subtracted flare spectra. The magnitude of the \textit{GOES} temperature
     and emission measure is indicated for each time interval of interest. }
   \label{fig:evegoes}
 \end{figure*}

 \begin{figure*}[t!]
   \centerline{\includegraphics[height=7.0in,angle=90]{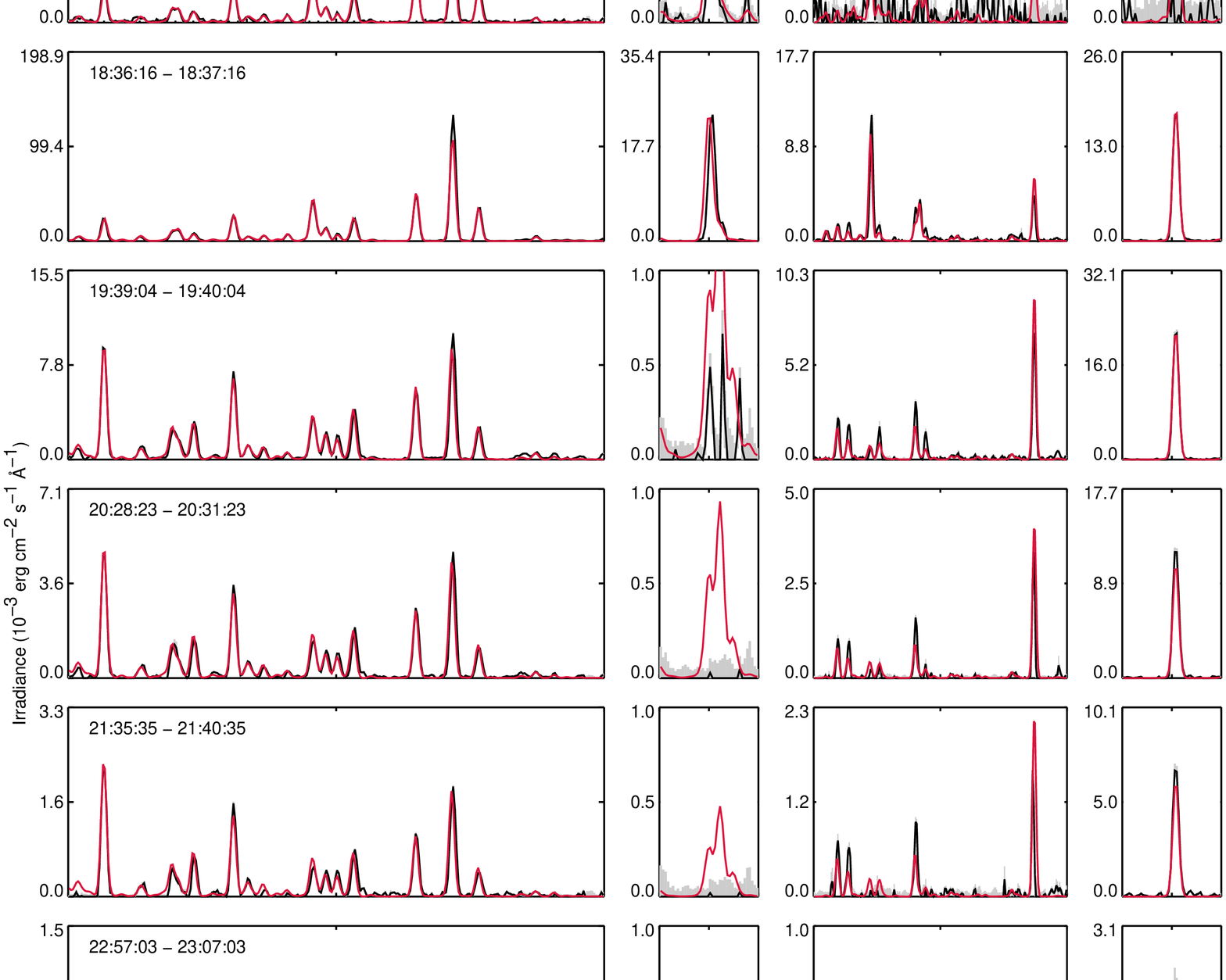}}
   \caption{Comparison of observed EVE spectra with spectra based on a DEM model. The DEMs
     are shown in Figure~\protect{\ref{fig:evedem2}}.}
   \label{fig:evedem1}
 \end{figure*}

 \subsection{Electron Densities}

 Most emission lines formed at flare temperatures are largely insensitive to the density
 and it is possible to compute the emission measure using any reasonable value. But, since
 observed intensity is very sensitive to the density, such measurements do provide an
 important constraint for hydrodynamic modeling. Anticipating this we consider the
 evolution of densities for this event.

 As mentioned in the introduction, there are several emission lines in the 90 to 150\,\AA\
 wavelength range from \ion{Fe}{21} that form density sensitive pairs
 \citep{mason1979}. More recent theoretical calculations are available from the CHIANTI
 atomic physics database \citep[e.g.,][]{dere1997,dere2009,landi2012}. The application of
 these ratios to the EVE data have been considered recently by \cite{milligan2012}. Their
 analysis suggests that the 145.73/128.75\,\AA\ ratio is the most useful. This ratio has a
 larger dynamic range than the (142.14+142.28)/128.75\,\AA\ ratio and is not as blended as
 121.21/128.75\,\AA. For very large events all three ratios yield similar
 results. Following \cite{milligan2012} we use the peak intensity above the local
 continuum as a proxy for the total intensity. As is shown in Figure~\ref{fig:eve2}, the
 145.73/128.75\,\AA\ ratio is approximately 0.02 over most of the event indicating an
 electron density of approximately $10^{11}$\,cm$^{-3}$. The peak temperature of formation
 of \ion{Fe}{21} is $10^7$\,K and we will assume a constant pressure of $10^{18}$\,K
 cm$^{-3}$ for these calculations.

 We note that because of the weakness of the \ion{Fe}{21} 145.73\,\AA\ lines, densities
 are only available near the peak of the event. The \textit{GOES} light curves and the AIA
 images show flare emission extending to at least 2\,UT on 2012 January 28. It is likely
 that the densities are lower during the decay and these measured values serve as an upper
 bound.

 \subsection{Isothermal Emission Measure}

 The \textit{GOES} observations can be used to compute an isothermal temperature and
 emission measure as a function of time. This calculation is based on theoretical spectra
 determined from version 7 of the CHIANTI atomic database that have been convolved with
 the spectral responses of the two channels. Additional details on the temperatures and
 emission measures derived from the \textit{GOES} observations are provided in
 \citep{white2005}. To isolate the emission from the flare, pre-flare fluxes are
 subtracted from the observed light curves. The temperature and emission measure for this
 event are displayed in Figure~\ref{fig:goes}. These calculations are performed using the
 IDL routine \verb+goes_chianti_tem+ distributed in the Solar Software Library (SSW,
 \citealt{freeland1998}).  As has been noted in previous analyses
 \citep[e.g.,][]{sterling1997} the highest temperatures are observed before the peak in
 the emission measure. Here we also see that the emission measure decays exponentially in
 time while the temperature is approximately constant during the decay of the event. A
 relatively constant temperature evolution has been noted in previous analysis (see, for
 example, \citealt{doschek1980}). We will return to this point in the next section.

 In converting from the temperature and volume emission measure derived from \textit{GOES}
 to the irradiance measured with EVE it is useful to recall that the irradiance is simply
 the intensity, or radiance, multiplied by the solid angle
 \begin{equation}
   I(\lambda) = \frac{A}{R^2}\left[
             \frac{1}{4\pi}\int \epsilon(\lambda,T_e,n_e)\xi(T_e)\,dT_e\right],
 \label{eq:flux}           
 \end{equation}
 where $A$ is the area of the feature, $R$ is the Earth-Sun distance, $A/R^2$ is the solid
 angle, and $\xi(T_e)=n_e^2ds/dT$ is the line of sight DEM. Note that the spatially
 unresolved \textit{GOES} observations yield a volume emission measure ($\xi_V=A\xi(T_e)$)
 which incorporates the area factor into the line-of-sight emission measure.  The
 isothermal \textit{GOES} model is equivalent to a $\delta$ function emission measure
 distribution
 \begin{equation}
   \xi_V(T_e) = \textrm{EM}_0\delta (T_e-T_0).
 \end{equation}

 To facilitate the rapid calculation of synthetic EVE spectra we have computed a grid of
 emissivities as a function of wavelength, temperature, and density
 ($\epsilon(\lambda,T_e,n_e)$) using the CHIANTI atomic database. We assume the CHIANTI
 ionization fractions and coronal abundances \citep{feldman1992}. We interpolate on this
 grid to produce a spectrum at a specified temperature and density. For this work we chose
 a spectral binning of 0.1\,\AA\ and convolve with a Gaussian smoothing function to
 account for the instrumental broadening. We find that a FWHM of 0.7\,\AA\ best reproduces
 the observed line widths in MEGS-A.
 
 In Figure~\ref{fig:evegoes} we compare the EVE spectra inferred from the \textit{GOES}
 model with the actual observations at various times during the event. Note that no
 scaling factors have been applied to either the spectra inferred from \textit{GOES} or
 the actual EVE observations. The wavelength range between 90 and 150\,\AA\ is generally
 well matched by the \textit{GOES} model. The most significant descrepancy is for the
 \ion{Fe}{18} emission at 93.93 and 103.95\,\AA. Similarly, the wavelength range near
 192\,\AA, which contains the \ion{Fe}{24} 192.04\,\AA\ line, is also well matched by the
 \textit{GOES} model.  The wavelength range between 245 and 290\,\AA\ contains a mixture
 of high temperature flare lines, such as \ion{Fe}{24} 255.10\,\AA\ and \ion{Fe}{23}
 263.76\,\AA, and lower temperature emission lines from \ion{Fe}{15} and \ion{Fe}{17}
 (see, for example, \citealt{warren2008} and \citealt{delzanna2008} for a description of
 the high temperature emission in this wavelength range). In this wavelength range the
 \textit{GOES} model only reproduces the highest temperature emission. Finally, we also
 see that the isothermal \textit{GOES} model does not reproduce any of the observed
 \ion{Fe}{16} 335.41\,\AA\ irradiance. The inability of the \textit{GOES} single
 temperature model to reproduce the observed EVE emission over a wide range of
 temperatures suggest that the flare is not isothermal.

 \begin{figure}[t!]
   \centerline{\includegraphics[height=3.15in,angle=90]{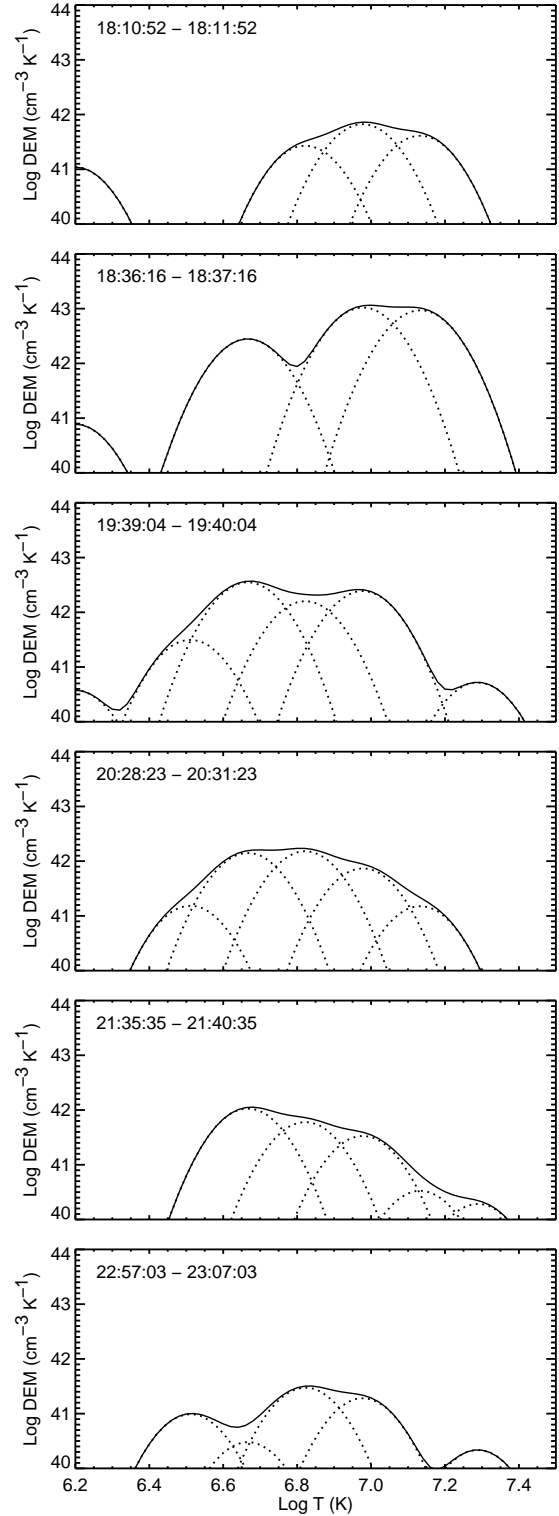}}
   \caption{DEM distributions computed from EVE spectra taken during the 2012 January 27
     event. The solid line is the actual distribution. The dotted lines indicate the
     contribution of each Gaussian component. }
   \label{fig:evedem2}
 \end{figure}

 \subsection{Differential Emission Measure}

 \begin{figure}
   \centerline{%
     \includegraphics[width=3.5in]{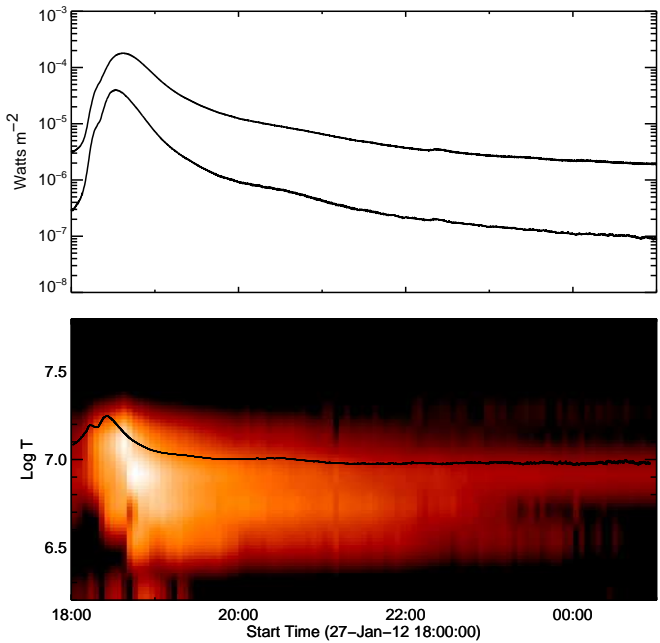}}
   \caption{(bottom panel) The DEM distribution as a function of temperature and time
     derived from the EVE data. Also shown is the isothermal \textit{GOES} temperature as
     a function of time. (top panel) The \textit{GOES} soft X-ray light curves for this
     time period.}
   \label{fig:dem_t1}
 \end{figure}

 \begin{figure*}[t!]
   \centerline{%
     \includegraphics[width=3.5in]{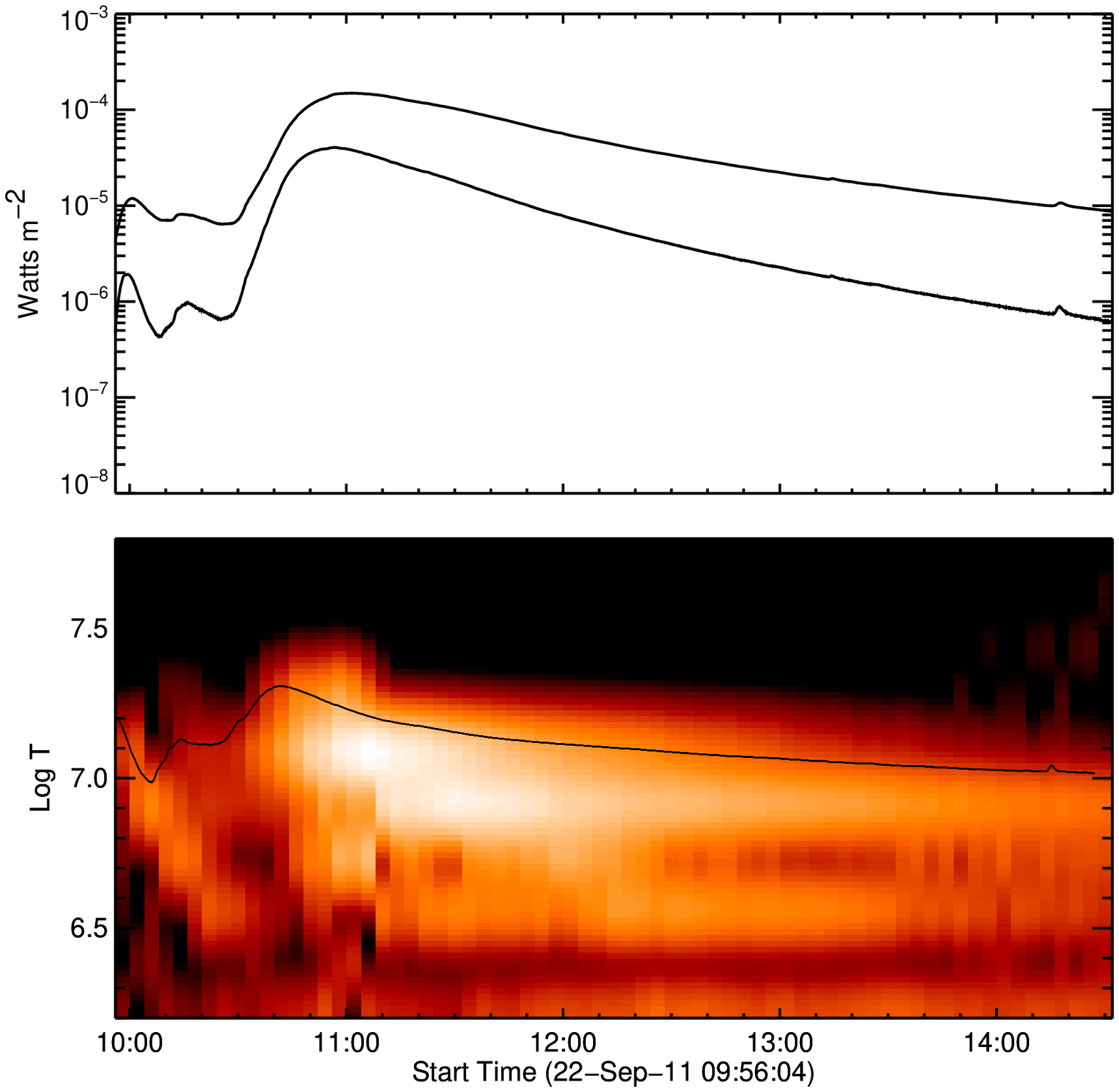}
     \includegraphics[width=3.5in]{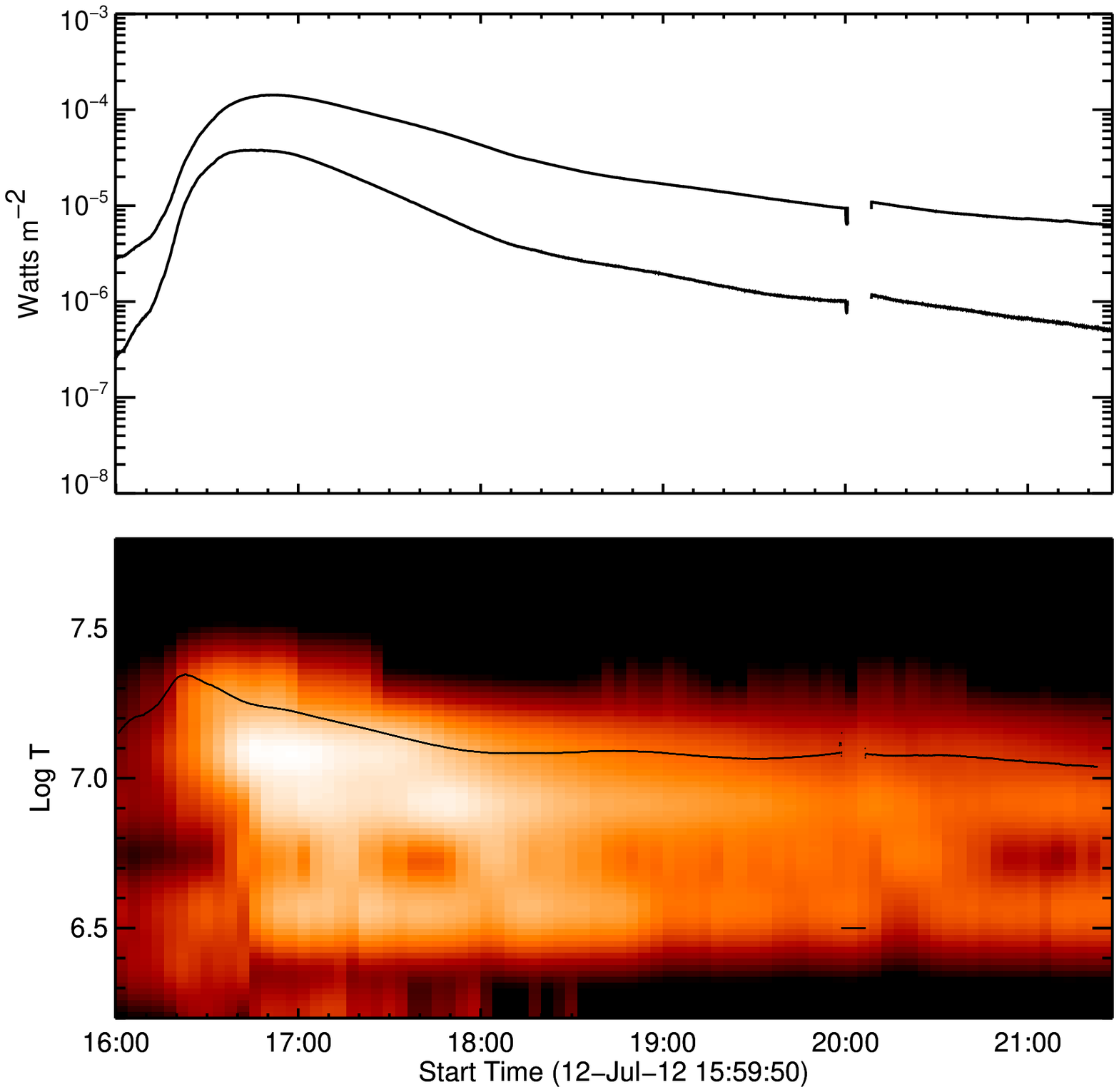}}
   \centerline{%
     \includegraphics[width=3.5in]{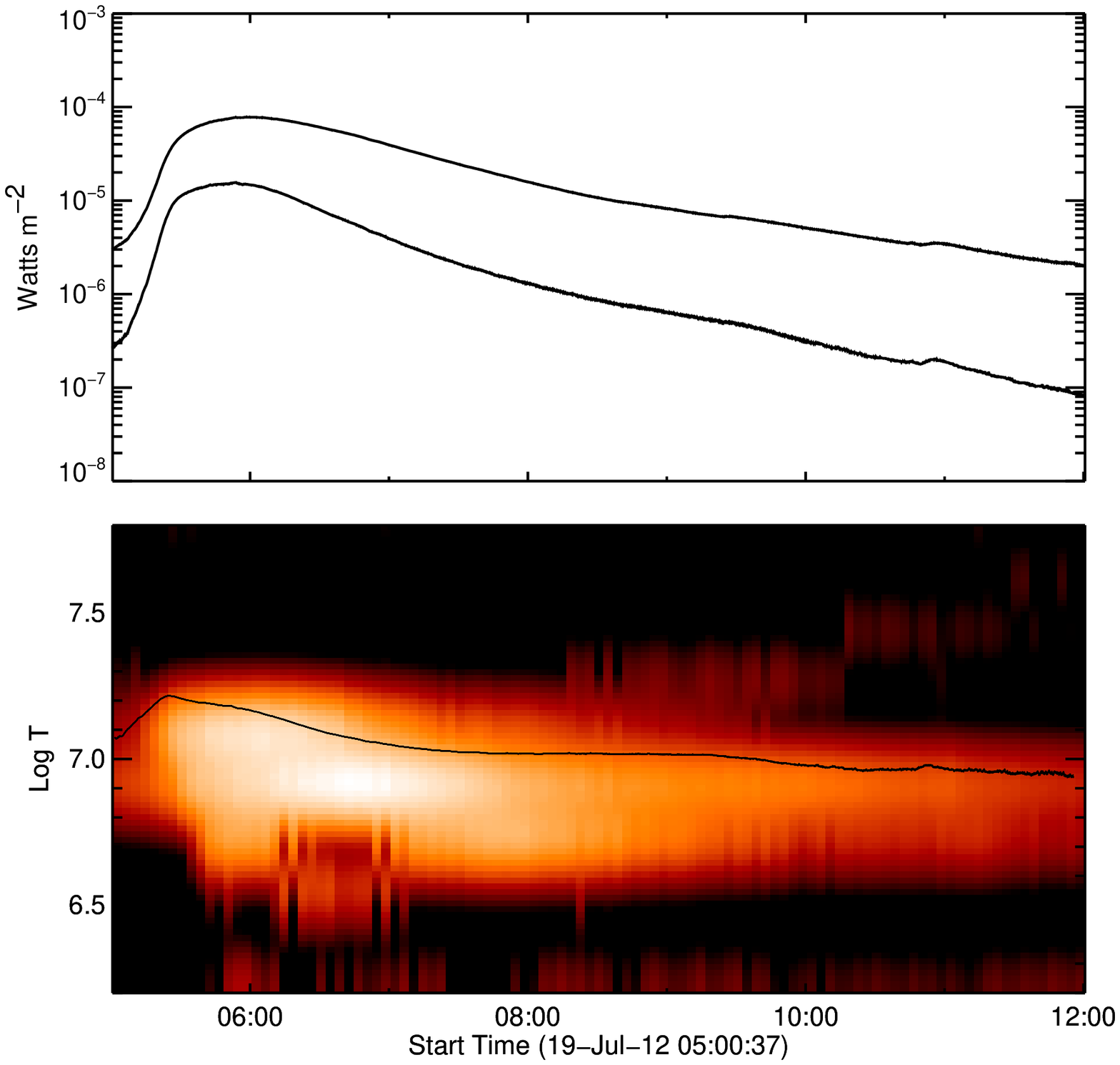}
     \includegraphics[width=3.5in]{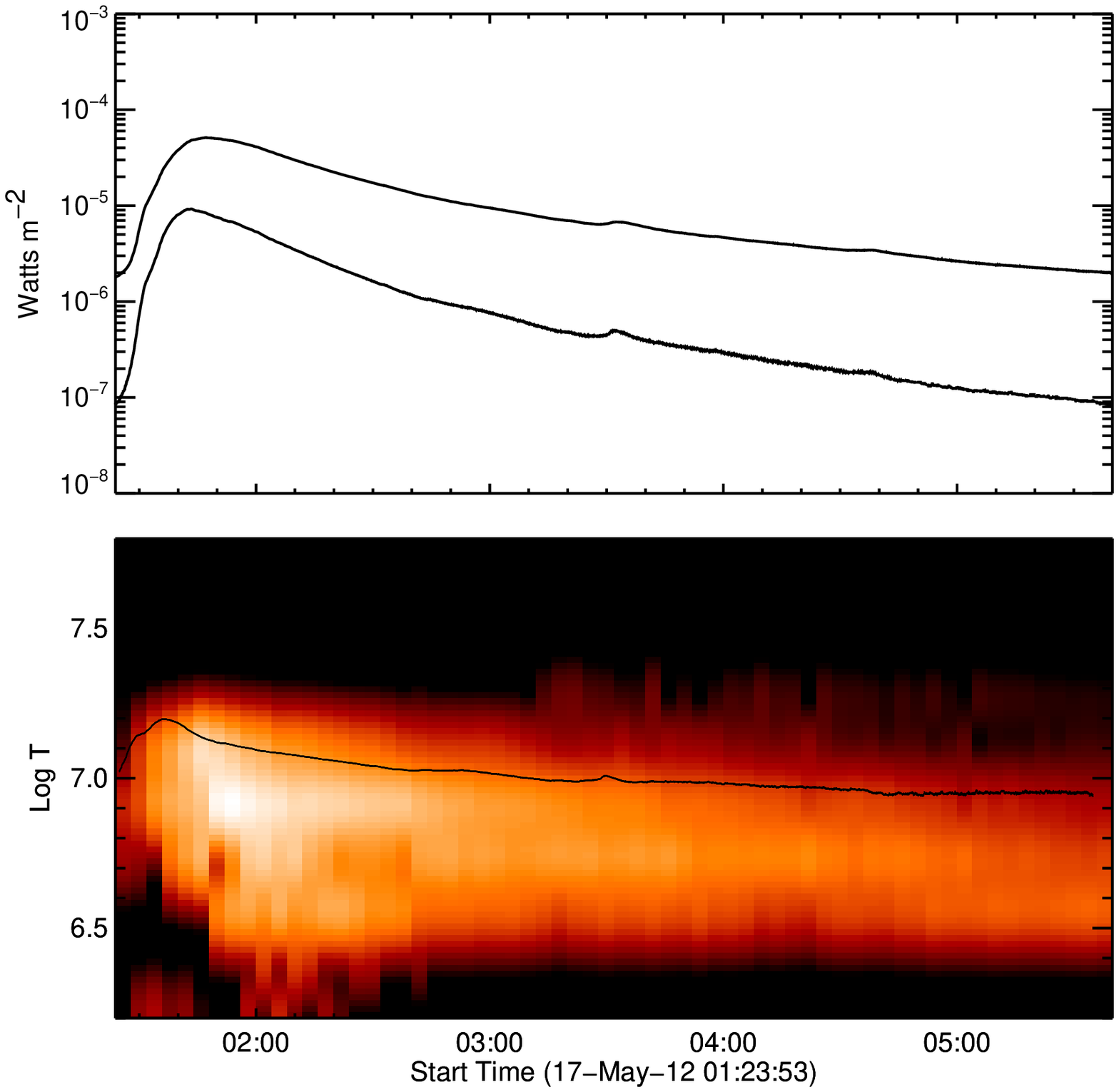}}
   \caption{Time-dependant DEM calculations for 4 long duration events associated with
     coronal mass ejections. The format for each plot is the same as
     Figure~\protect{\ref{fig:dem_t1}}.}
   \label{fig:dem_t2}
 \end{figure*}

 To solve for the temperature distribution implied by the observed spectra we assume that
 the volume DEM can be approximated as a sum of Gaussians in log space,
 \begin{equation}
   \xi_V(T_e) = \sum_{k=1}^{N_g} \textrm{EM}_k \exp\left[
      -\frac{(\log T_e-\log T_k)^2}{2\sigma_k^2}\right],
 \end{equation}
 where the number ($N_g$), position ($\log T_k$), and width ($\sigma_k$) of the Gaussians
 is fixed for a given calculation and only the magnitude of each component is varied. In
 general it is not possible to obtain well behaved solutions to integral equations such as
 Equation~\ref{eq:flux} \citep[e.g.,][]{craig1976}. Solutions are generally sensitive to
 noise and the assumption of multiple Gaussians represents an attempt to regularize or
 smooth the DEM while maintaining the ability the reproduce relatively isothermal
 distributions.

 To solve for $\xi_V(T)$ we define a temperature domain of interest $[T_1,T_2]$ and select
 a value for $N_g$. We assume that components are equally spaced and that the width of the
 component is related to the width of the temperature domain by
 \begin{equation}
 \sigma_k = \sigma = \frac{\log T_2-\log T_1}{2N_g}.
 \end{equation}
 We then select initial values for $\textrm{EM}_k$ and use the Levenberg-Marquardt
 least-squares minimization routine \verb+MPFIT+ to determine the values that produce the
 lowest value of $\chi^2$. In computing the differences between the spectrum computed
 using the emission measure model and the observed spectrum we only consider the
 wavelength ranges shown in Figures~\ref{fig:evegoes} so that lines at other wavelengths,
 such as \ion{He}{2} 304\,\AA, do not affect the calculation.

 In Figures~\ref{fig:evedem1} and \ref{fig:evedem2} we show the EVE spectra inferred from
 the DEM inversion and the individual DEMs for several times during the event. These
 calculations were preformed assuming $N_g=10$. We have investigated how the goodness of
 fit depends on the assumed number of components and found that the $\chi^2$ declines only
 by a small amount for values of $N_g$ below approximately 8 to 12. We also varied the
 assumed instrumental FWHM and found that a value of 0.7\,\AA\ produces the lowest
 $\chi^2$. Finally, we also confirmed that using different values for the assumed pressure
 has very little influence on the goodness of fit.

 For all of the spectra considered here the DEM model reproduces the observations more
 closely than the isothermal model does. The isothermal model produces $\chi^2$ values
 that are 5--10 times higher than the DEM model, although this is due in part to absolute
 differences in calibration between EVE and \textit{GOES}. The DEM model accounts for the
 cooler \ion{Fe}{15}, \ion{Fe}{16}, and \ion{Fe}{18} emission lines in the observed
 spectra. The discrepancies between the computed and observed spectra for the higher
 temperature emission lines are also reduced.  For the 19:39, 20:28, and 21:35 spectra,
 however, the DEM model appears to overestimate the observed emission near 192\,\AA. Since
 this discrepancy is actually smaller for the \textit{GOES} isothermal spectra, it
 suggests a sharp drop in the DEM above a temperature of approximately $10^7$\,K. Adding
 additional Gaussian components to the DEM, however, does not reduce the intensity of the
 \ion{Fe}{24} 192.04\,\AA\ line significantly. Since this region of the spectrum is
 dominated by emission lines formed near 1\,MK it seems likely that there actually is some
 \ion{Fe}{24} emission during this time, it is just difficult to observe in the pre-flare
 subtracted spectra. The absence of detectable emission from \ion{Ca}{17} 192.86\,\AA,
 which is formed at about $\log T_e = 6.5$, is consistent with this interpretation. We
 also note that the discrepancy between the observed spectra and the DEM model spectra for
 \ion{Fe}{24} 255.10\,\AA\ line is generally small.

 The selection of a limited number of intervals for considering the DEM allows us to make
 detailed comparisons between the modeled and observed spectra as well as with parameters
 inferred from the GOES observations. The EVE data, however, are available at a cadence of
 10\,s allowing the temperature evolution to be followed at high cadence over long periods
 of time. The isothermal temperature derived from \textit{GOES} shown in
 Figure~\ref{fig:goes} suggests that analysis of high cadence data would be of interest.
 The isothermal model indicates a rapid increase in the temperature during the rise phase
 of the flare followed by a decline to a relatively constant temperature of approximately
 10\,MK during the decay. The DEMs shown in Figure~\ref{fig:evedem2} appear to be
 consistent with this behavior. The DEMs from 18:10 and 18:36 show considerable emission
 at temperatures above $\log T_e = 7.1$.  At later times most of the emission lies at
 temperatures between $\log T_e = 6.5$ and 7.1.

 To compute the DEM as a function of temperature and time we divide the EVE observations
 during the flare into 120\,s intervals and compute the DEM for each interval as described
 previously. In Figure~\ref{fig:dem_t1} we show the result of this calculation. The DEM
 during the rise of the flare is dominated by very high temperature plasma. Near the peak
 of the flare the DEM becomes very broad with strong emission at temperatures between
 $\log T_e = 6.4$ and 7.4. During the decay the highest temperature emission fades away as
 the DEM assumes an approximately constant shape. At the very end of the event the DEM at
 the lowest temperatures also becomes small. It seems likely, however, that this is due to
 the difficulty of separating the flare from the background irradiance in \ion{Fe}{15}
 284.16\,\AA\ and \ion{Fe}{16} 335.41\,\AA. As is seen in Figure~\ref{fig:aia}, the AIA
 images from this time continue to show the formation of relatively weak post-flare loops
 at a wide range of temperatures.

 Comparisons between the EVE DEM and the isothermal \textit{GOES} temperature show that
 the evolution of the \textit{GOES} temperature is consistent with the evolution of the
 DEM. These comparisons also show that the \textit{GOES} temperature is strongly weighted
 towards the highest temperatures in the flare. 

 We have preformed this time-dependent DEM calculation on 4 other long duration events
 that were associated with coronal mass ejections observed in AIA.  The DEMs for these
 events are shown in Figure~\ref{fig:dem_t2} and are generally similar to those computed
 for the 2012 January 27 event. All of the events that we have studied show a broad
 distribution of temperatures throughout the entire evolution of the flare. During the
 rise phase and at the peak of the events we find the highest temperatures and a rapid
 evolution in the DEM. During the decay we find somewhat lower peak temperatures and an
 approximately constant shape for the DEM.
 
 \section{Discussion}

 The combination of continuous observations, broad wavelength coverage, and relatively
 high spectral resolution of the EVE instrument on \textit{SDO} provide a new opportunity
 to study the evolution of thermal flare plasma in detail. We have shown that these
 observations can be used to construct DEMs between approximately $\log T_e = 6.3$ and
 7.5. Flare related emission at lower temperatures is clearly evident in the AIA images,
 but it is difficult to isolate this signal in the spatially integrated irradiance
 observations. The highest temperature emission lines observed by EVE are from
 \ion{Fe}{24}, which limits the DEM at the highest temperature. EVE is unlikely to be able
 to detect low emission measure, ``super hot'' plasma ($\log T_e > 30$\,MK; see, for
 example \citealt{caspi2010}). It should be possible, however, to combine the EVE data
 with high energy observations from RHESSI to provide a more complete description of the
 thermal and non-thermal emission in flares.

 EVE measurements of thermal flare plasma evolution provide important constraints on
 theories of energy release during a flare. These observations also provide useful tests
 on the hydrodynamics of loop evolution.  For the events considered here it is clear from
 both the broad temperature distributions and the AIA images showing emission over a wide
 range of temperatures that flares are not consistent with the evolution of a small number
 of loops. A more likely scenario is the continuous formation of loops that are initially
 heated to high temperature and then cool. This idea of heating and cooling occurring on a
 succession of independently heated loops has been incorporated into simple hydrodynamic
 flare models that can reproduce not only the evolution of the observed intensities
 \citep{hori1997,hori1998,reeves2002,warren2006,reeves2010} but also detailed properties
 of the line profile \citep{warren2005}. Most of this work, however, has focused on the
 emission at the highest temperatures as well as on the evolution of the thermal plasma
 during the rise and peak of an event. It remains to be seen if these models can reproduce
 the distribution of temperatures observed by EVE over the full evolution of a flare.

 As noted in the introduction, the evolution during the extended decay of a flare holds
 particular promise for probing the fundamental process of magnetic reconnection. During
 this time we see a relatively constant shape to the emission measure over many hours. The
 emission measure, however, is decaying exponentially, indicating that the temperature and
 the density are reacting very differently to changes in the heating rate. Simple
 hydrodynamic arguments by \cite{warren2004} have shown that the peak density of an
 impulsively heated loop scales as
 \begin{equation}
   n\sim\left(\frac{E}{A}\right)^{2/3}\frac{1}{L}
 \end{equation}
 while the temperature at the time of the peak density scales as
 \begin{equation}
   T\sim\left(\frac{E}{A}\right)^{1/3},
 \end{equation}
 where $E$ is the total energy input, $A$ is the cross-sectional area, and $L$ is the loop
 length.  For the two ribbon events considered here we anticipate that the broad
 temperatures distributions we measure will require that the flares be modeled as a
 succession of impulsively heated loops. Furthermore, we anticipate that over time the
 energy input into each loop will decline. These relationships suggest that the decline in
 input energy for each newly formed loop will lead to relatively large changes in the
 magnitude of the emission measure over time while leaving the temperature structure
 relatively unchanged. It remains to be demonstrated, however, that simple hydrodynamic
 models can reproduce the EVE observations in detail.


 \acknowledgments The \textit{SDO} mission and this research was supported by NASA. HPW
 thanks Amir Caspi and Jim McTiernan for many interesting discussion on EVE flare
 observations.


\bibliography{apj}

\end{document}